\documentclass[twocolumn,preprintnumbers,amsmath,amssymb,prd]{revtex4}

\usepackage{hyperref}
\usepackage{graphicx}
\usepackage{dcolumn}
\usepackage{bm}

\begin{document}

\preprint{gr-qc/0409073}

\title{Hyperboloidal Slices and Artificial Cosmology for Numerical Relativity}

\author{Charles W.\ Misner}
\email{misner@physics.umd.edu}
\affiliation{Department of Physics, University of Maryland,
College Park, MD 20742-4111, USA}

             
\date{21 January 2005}

\begin{abstract}
This preliminary report proposes integrating the Maxwell equations in
Minkowski spacetime using coordinates where the spacelike surfaces are
hyperboloids asymptotic to null cones at spatial infinity. The space
coordinates are chosen so that Scri+ occurs at a finite coordinate and a
smooth extension beyond Scri+ is obtained. The question addressed is
whether a Cauchy evolution numerical integration program can be easily
modified to compute this evolution.
  In the spirit of the von Neumann and Richtmyer artificial viscosity
which thickens a shock by many orders of magnitude to facilitate
numerical simulation, I propose artificial cosmology to thicken null
infinity Scri+ to approximate it by a de Sitter cosmological horizon
where, in conformally compactified presentation, it provides a shell of
purely outgoing null cones where asymptotic waves can be read off as
data on a spacelike pure outflow outer boundary. This should be simpler
than finding Scri+ as an isolated null boundary or imposing outgoing
wave conditions at a timelike boundary at finite radius.
\end{abstract}

\maketitle

{\em It is a pleasure to dedicate this work to my friend and colleague
Stanley Deser as we celebrate his ancient works and his continuing
contributions to a broad variety of problems in physics.}

\section{\label{sec:Intro}Introduction}
As LIGO \cite{LIGO} approaches its first stage design sensitivity and
other \cite{VIGO} gravitational wave observatories progress, theoretical
descriptions of the expected waveforms from the inspiral of binary black
holes and neutron stars remain less detailed than one might hope. The
largest efforts are devoted to solving Einstein's equations by discrete
numerical methods based on a space+time split of the equations and
intend to impose as a boundary condition an asymptotic Minkowski metric.
Initial formulations were given the ADM name although (probably
inconsequentially) none uses the conjugate sets of fields $g_{ij}$ and
$\pi^{ij}$ favored by ADM. Variations on these formulations have shown
practical improvements \cite{bs:bssn,sn:bssn,aei:bssn}, 
and others \cite{ls:sh} with less developed implementations
hold the theoretical suggestion of better numerical behavior.

Because the (expected) observable gravitational waves will be seen far
from their sources, some efforts have asked that the data set
representing the state of the gravitational field at any computational
step be not that on an asypmtotically flat spacelike hypersurface, but
instead that on an outgoing light cone. Thus the Pittsburgh group \cite{win:lrr} has
focussed on slicing spacetime along null cones. By extracting a
conformal factor as in Penrose diagrams, null infinity becomes a finite
boundary in the computational grid. But these outgoing null cones do not
have a useful behavior near the wave sources, leading to a study
matching them to flatter spacelike slices there. Friedrich \cite{hf:sh} formulated
the Einstein equations in a way where the conformal factor could be one
of the dynamical variables, and also considered the case where the time
slices were spacelike but asymptotically null, as for hyperboloids in
Minkowski spacetime. A large program to implement this approach is being
pursued by Sascha Husa \cite{sascha:tuebingen,sascha:group} and others at AEI-Potsdam. 

This paper aims to formulate within the simpler problem of solving
Maxwell's equations some aspects of these methods which could be
attacked with smaller computational resources. In particular we consider
the question of whether fairly straightforward evolution algorithms
designed for spacelike slicings of spacetime will encounter special
problems when those slices are asymptotically null, and the question of
imposing boundary conditions at null infinity (Scri+).

Work with Mark Scheel and Lee Lindblom \cite{MSL:KITP} considered the
first of these questions, but employed the scalar wave equation. Those
preliminary results were presented at the KITP Workshop ``Gravity03'' in
June 2003. Vince Moncrief \cite{VMpc} then suggested that since the coordinate
choice which brought Scri+ to a finite radius displayed a conformal
metric which was nonsingular at Scri+, it would be better to study
instead the Maxwell equations whose properties under conformal
transformations are simpler that those of the scalar wave equation.

A new suggestion is given below that the boundary conditions at Scri+
might be easier to handle if Scri+ were approximated by a de Sitter
cosmological horizon with a cosmological constant chosen for numerical
convenience rather than for physical interest.

\section{\label{sec:MaxEq}Maxwell's Equations}

In formulating a 3+1 statement of the Maxwell equations I, of course,
revert to lessons from Arnowitt and Deser \cite{AD59}. Following their
Schwinger tradition I use a first order variational principle $\delta I
= 0$ with
\begin{equation}
       I = \int g^{\mu\nu\,\alpha\beta}(-F_{\mu\nu} \partial_\alpha A_\beta
           + \tfrac{1}{4} F_{\mu\nu} F_{\alpha\beta})\,d^4x
\label{eq:MaxI}
\end{equation}
where
\begin{equation}
       g^{\mu\nu\,\alpha\beta} = \tfrac{1}{2}\sqrt{-g}
                        (g^{\mu\alpha} g^{\nu\beta}
                      - g^{\mu\beta} g^{\nu\alpha})
\label{eq:gup4}
\end{equation}
are components of an operator that maps covariant anti-symmetric tensors
onto contravariant anti-symmetric tensor densities.
This mapping has an inverse whose components are
\begin{equation}
       g_{\mu\nu\,\alpha\beta} = \tfrac{1}{2}
                        (g_{\mu\alpha} g_{\nu\beta}
                      - g_{\mu\beta} g_{\nu\alpha})/\sqrt{-g} \quad .
\label{eq:gdn4}
\end{equation}
Note that these two symbols are {\em not} related merely by the usual
lowering of indices, as one map raises the density while the other
lowers it.
Their product
\begin{equation}
       g_{\mu\nu\,\rho\sigma} g^{\rho\sigma\,\alpha\beta} = 
           \tfrac{1}{2}(\delta_\mu^\alpha \delta_\nu^\beta
                      - \delta_\mu^\beta \delta_\nu^\alpha) \quad .
\label{eq:id4}
\end{equation}
gives the identity mapping of antisymmetric tensors onto themselves.

In this variational principle the fields  $A_\mu$ and  $F_{\mu\nu}$
are varied independently and yield the equations
\begin{equation}
       \partial_\nu \mathfrak{F}^{\mu\nu} = 0
\label{eq:MaxDiv}
\end{equation}
and
\begin{equation}
       F_{\mu\nu} = \partial_\mu A_\nu - \partial_\nu A_\mu \quad .
\label{eq:MaxCurl}
\end{equation}
Here $\mathfrak{F}^{\mu\nu} = \sqrt{-g}F^{\mu\nu}$ arises as
\begin{equation}
       \mathfrak{F}^{\mu\nu} = g^{\mu\nu\,\alpha\beta} F_{\alpha\beta}
            \quad .
\label{eq:frakF}
\end{equation}

Again following the given tradition I seek a proto-Hamiltonian form $p
\dot{q}-H$ for this Lagrangian and rewrite it by replacing the
$F_{\mu\nu}$ field in $I$ by its equivalent $\mathfrak{F}^{\mu\nu}$
field, giving
\begin{equation}
       I = \int (-\mathfrak{F}^{\mu\nu} \partial_\mu A_\nu
           + \tfrac{1}{4} \mathfrak{F}^{\mu\nu} \mathfrak{F}^{\alpha\beta} 
              g_{\mu\nu\,\alpha\beta})\,d^4x \quad .
\label{eq:MaxI3}
\end{equation}
The only term here containing time derivatives is $-\mathfrak{F}^{0i}
\partial_0 A_i$ so that 
\begin{equation}
       \mathfrak{D}^i \equiv \mathfrak{F}^{0i} 
\label{eq:Edef}
\end{equation}
and $A_i$ are the only fields for which one obtains evolution equations.
The scalar potential
\begin{equation}
       \psi \equiv -A_0
\label{eq:psidef}
\end{equation}
and the $\mathfrak{F}^{ij}$ are therefore Lagrange multipliers which 
enforce the constraints
\begin{subequations}
\label{eq:constr}
\begin{equation}
\label{eq:Econstr}
  \partial_i \mathfrak{D}^i = 0
\end{equation}
\text{and}
\begin{equation}
\label{eq:Bconstr}
     B_{ij} \equiv g_{ij,\alpha\beta}  \mathfrak{F}^{\alpha\beta} 
         = \partial_i A_j - \partial_j A_i \quad .
\end{equation}
\end{subequations}
The evolution equations can then be written as
\begin{subequations}
\label{eq:evol}
\begin{equation}
\label{eq:Aevol}
  \partial_0 A_i = -E_i - \partial_i \psi  
\end{equation}
\text{and}
\begin{equation}
\label{eq:Eevol}
   \partial_0 \mathfrak{D}^i = \partial_j \mathfrak{H}^{ij}
\end{equation}
\end{subequations}
where
\begin{equation}
\label{eq:EHdef}
    E_i \equiv g_{i0,\alpha\beta}  \mathfrak{F}^{\alpha\beta}
              \quad\text{and}\quad
    \mathfrak{H}^{ij} \equiv  \mathfrak{F}^{ij} \quad .
\end{equation}

There is no evolution equation for $\psi = -A_0$ but one can be supplied
as a gauge condition. There is also no evolution equation directly from
the variational principle for $B_{ij}$ but one can be deduced by taking
the time derivative of equation (\ref{eq:Bconstr}) and evaluating the 
time derivatives on the right hand side using equation (\ref{eq:Aevol}).
The result is
\begin{equation}
       -\partial_0 B_{ij}  =  \partial_i E_j - \partial_j E_i \quad .
\label{eq:Bevol}
\end{equation}

In a numerical example mentioned below, the fields were taken to be
$A_i$ and $\mathfrak{D}^i$ using the evolution equations
(\ref{eq:Aevol}) and (\ref{eq:Eevol}). Equation (\ref{eq:Bconstr}) was
treated not as a constraint but as merely making $B_{ij}$ an
abbreviation for the right member of that equation, thus introducing
second (spatial) derivatives into the system of equations. For that
example of a wave packet the gauge chosen was Coulomb gauge with
$\psi=0$ for all time. The remaining constraint (\ref{eq:Econstr}) is
easily seen to be preserved in time as a consequence of the evolution
equation (\ref{eq:Eevol}).

Another possible system of equations would be to use (\ref{eq:Eevol})
and (\ref{eq:Bevol}) as evolution equations. Since the vector potential
does not appear in these equations, it can be ignored and not evolved.
The constraint (\ref{eq:Bconstr}) would then have to be replaced by its
integrability condition 
\begin{equation}
\label{eq:dBconstr}
     \partial_{[k} B_{ij]} = 0 
\end{equation}
treated as a constraint equation. It is easily seen from
(\ref{eq:Bevol}) that this constraint is preserved in time
(differentiably) as a consequence of (\ref{eq:Bevol}). The Gauss
constraint (\ref{eq:Econstr}) remains part of this system and is
preserved in time by exact solutions. 
These equations (\ref{eq:Eevol})
and (\ref{eq:Bevol}) are {\em exactly} what one writes in classical
electromagnetism in a material medium (e.g., \cite[Section
16-2]{ReitzMC:elmag}); only the constitutive equations (\ref{eq:defEH})
are new. 
With the usual 3+1 decomposition of the metric
\begin{equation}
       ds^2 = - \alpha^2 dt^2 
              + \gamma_{ij}(dx^i + \beta^i du) (dx^j + \beta^j du)
\label{eq:g3+1}
\end{equation}
these are
\begin{subequations}
\label{eq:defEH}
\begin{equation}
\label{eq:defE}
  E_i = (\alpha/\sqrt{\gamma})\gamma_{ij}\mathfrak{D}^j
                   + B_{ij} \beta^j 
\end{equation}
\text{and}
\begin{equation}
\label{eq:defH}
   \mathfrak{H}^{ij} =   \alpha \sqrt{\gamma}\gamma^{ik}\gamma^{jl}B_{kl}
                      +\mathfrak{D}^i \beta^j - \mathfrak{D}^j \beta^i 
\end{equation}
\end{subequations}
which are more complex than the $\bf{E} = \bf{D}/\epsilon$ and
$\bf{H} = \bf{B}/\mu$ of simple isotropic media. The Einstein \ae ther
is, in coordinate terms, anisotropic not only in space but also in
spacetime, but conveniently linear.

\section{\label{sec:conformal}Conformal Invariance}
This exploration of numerical evolution with the Maxwell equations was
provoked by Vince Moncrief's reminder that the Maxwell and Yang-Mills
equations (in four spacetime dimensions) have the simplest possible
conformal structure. A conformal transformation $g_{\mu\nu} \mapsto
\Omega^2 g_{\mu\nu}$ changes the metric geometry to a different one
which shares the same light cones. Under this change the metric
dependent factor $g^{\mu\nu\,\alpha\beta}$ in the variational integral
$I$ does {\em not} change since $g^{\mu\nu} \mapsto \Omega^{-2}
g^{\mu\nu}$ while $\sqrt{-g} \mapsto \Omega^4 \sqrt{-g}$. As nothing in
the variational integral changes under a conformal transformation, it
follows that if the fields $A_\mu$ and $F_{\mu\nu}$ or
$\mathfrak{F}^{\mu\nu}$ satisfy the field equations in one metric, so do
they also in any conformally related metric.

Although nothing in section~\ref{sec:MaxEq} depends on the metric having
any special properties (other than Lorentz signature) such as flatness
or satisfying the Einstein equations, these conformal properties in four
spacetime dimension are particularly useful since the metric
(\ref{eq:deSitterH}) we intend to use is singular only in a conformal
factor $s^2/q^2$ which we now see will appear nowhere in our field
equations. A similar simplification occurs when the metric of interest
is the Schwarzschild metric in the hyperboloidal slicings used in
\cite{MSL:KITP}.

In the 3+1 decomposition of the field equations we note that $A_i,
\mathfrak{D}^i, E_i, \psi, B_{ij}, \mathfrak{H}^{ij}$ and $\beta^i$ are
conformally invariant, while $\gamma_{ij} \mapsto \Omega^2 \gamma_{ij},
\alpha \mapsto \Omega \alpha, \sqrt{\gamma} \mapsto \Omega^3
\sqrt{\gamma}$. Thus the metric structures which appear in the
constituitive equations (\ref{eq:defEH}),
$(\alpha/\sqrt{\gamma})\gamma_{ij}$ and $\alpha
\sqrt{\gamma}\gamma^{ik}\gamma^{jl}$ as well as $\beta^i$, are
conformally invariant.

The constraint equations (\ref{eq:constr}) and (\ref{eq:dBconstr}) have
an even stronger invariance: they are metric invariant. No metric
quantities appear in these constraint equations when they are written in
terms of our choice of fields. Thus if two hypersurfaces are
diffeomorphic then initial conditions for the Maxwell equations which
satisfy the constraints in one Lorentzian manifold can be imported to
the other where they will again satisfy the constraints. If the two
manifolds are not conformally equivalent, howeve, the subsequent
evolutions of these initial conditions will generally be inequivalent.

Gauge conditions present greater difficulties. In the absence of sources
the ``Coulomb'' gauge $\psi=0$ is conformally invariant, but only
3-dimensionally coordinate invariant. On the other hand the simplest
Lorentz gauge $\partial_\mu (\sqrt{-g} g^{\mu\nu} A_\nu) = 0$ is
coordinate invariant but not conformally invariant. The set of field
equations which evolve $\mathfrak{D}^i$ and $B_{ij}$ without the use of
the vector potential are conformally invariant but do not require a
gauge condition.

\section{\label{sec:deSitterMetric}Background Spacetime}
The background metric for this study will be the de Sitter spacetime
\begin{equation} 
ds^2 = -dT^2 +dX^2 + dY^2 + dZ^2 + (R^2/L^2)(dT-dR)^2 \quad . 
\label{eq:deSorig} 
\end{equation} 
which is here presented in Kerr-Schild form as flat spacetime plus the
square of a null form. The cosmological constant here is $\Lambda =
+3/L^2$. Since $dT-dR$ is null in both the underlying flat spacetime and
after the Kerr-Schild modification, the hyperboloidal slices we used in
flat spacetime should remain asymptotically null also in this modified
metric. Therefore we introduce the same coordinate change as in our
earlier scalar work \cite{MSL:KITP} 
\begin{equation}
  \frac{T}{s} =  u + \frac{\tfrac{1}{2}r^2}{1-\tfrac{1}{4}r^2}
\label{eq:HsliceU}
\end{equation}
and
\begin{equation}
  \frac{X^i}{s} = \frac{x^i}{1-\tfrac{1}{4} r^2} \quad  .
\label{eq:AnMR}
\end{equation}
where $R^2 = X^2 +Y^2 +Z^2 \equiv X^i X^i$ and $r^2 = x^2+y^2+z^2 \equiv
x^i x^i$.
This  leads to the metric
\begin{subequations}
\label{eq:deSitterH}
\begin{eqnarray}
  ds^2 & = & (s^2/q^2)[-\alpha^2 du^2   \\ 
       &   & + \left[1+\frac{s^2 r^2}{L^2 (1+\tfrac{1}{2}r)^4}\right]
            (dr +\beta du)^2  + r^2 d\Omega^2]  \nonumber
\end{eqnarray}
\text{where}
\begin{equation}
     q = 1 - \tfrac{1}{4} r^2 \quad ,
\end{equation}
\begin{equation}
      \alpha^2 = (1 + \tfrac{1}{4} r^2)^2
         \left[1+\frac{s^2 r^2}{L^2 (1 + \tfrac{1}{2} r)^4}\right]^{-1}
 \quad ,
\end{equation}
\text{and}
\begin{equation}
     \beta = -\left[r+\frac{s^2 r^2}{L^2 (1 + \tfrac{1}{2} r)^2}\right]
       \left[1+\frac{s^2 r^2}{L^2 (1 + \tfrac{1}{2} r)^4}\right]^{-1}
\end{equation}
\end{subequations}

I have verified that this metric (\ref{eq:deSitterH}) actually is the
de Sitter metric by two tests: I have run a GRtensor II \cite{grt} Maple
8 worksheet to calculate the Einstein equations for it and find
\begin{equation}
     \label{eq:EinEqsDeS}
     G^\mu_\nu =-(3/L^2) \delta^\mu_\nu
\end{equation}
and I have also had that worksheet calculate the Riemann tensor which is
\begin{equation}
     \label{eq:ReimannDeS}
     {R^{\mu\nu}}_{\alpha\beta} = (1/L^2)
   (\delta^\mu_\alpha\delta^\nu_\beta - \delta^\mu_\beta\delta^\nu_\alpha)
\end{equation}
and shows that this is a spacetime of constant curvature, therefore de
Sitter. Richard Woodard \cite{RWpc} has confirmed this by showing that
the metric (\ref{eq:deSorig}) can be reduced to the familiar static
deSitter metric by the coordinate change $T = \bar{T} + f(R)$ where
$df/dR = -R^2/(L^2 - R^2)$.

\section{\label{sec:Causal}Causal Structure}
This de Sitter metric is presented in equations (\ref{eq:deSitterH}) with
an apparent singularity at $r=2$ where $q=0$. This is probably a
coordinate singularity since the curvature does not go bad there, and
also because the de Sitter spacetime in notorious for being presented in
many different guises corresponding to different coordinate systems,
many of which cover only a patch of the full manifold. But for present
purposes we only need this coordinate patch, and the singularity at
$R=\infty$ will be removed by a conformal transformation later.

By dropping the conformal factor in the metric (\ref{eq:deSitterH}) we
arrive at a nonsingular metric
\begin{eqnarray}
  \label{eq:deSreg}
  ds^2 & = & -\alpha^2 du^2   \\ 
       &   & + \left[1+\frac{s^2 r^2}{L^2 (1+\tfrac{1}{2}r)^4}\right]
            (dr +\beta du)^2  + r^2 d\Omega^2  \nonumber
\end{eqnarray}
using the same abbreviations as in equations (\ref{eq:deSitterH}). Note,
however, that $\alpha$ and $\beta$ are now the lapse and shift of this
regulated metric. This metric no longer satisfies the Einstein equations
(with $\Lambda$ term) but is related to the original (de Sitter) metric
in a known way. From this point on we use only this metric, and the
previous metric will be called $d\tilde{s}^2 = \Omega^2 ds^2$ following
the practice in Wald's discussion \cite[Appendix D]{wald:wald} of
conformal transformations, with $\Omega = s/q$ in our case. 

\begin{figure}
\includegraphics[scale=0.3, angle=-90]{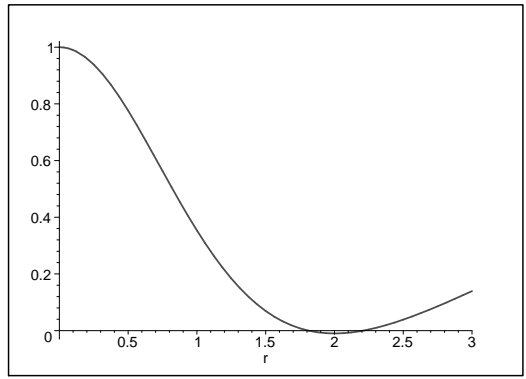}%
\caption{\label{fig:gradr2macro}
The squared norm $(\nabla r)^2 = g^{rr}$ of the normal $\nabla r$ to
hypersurfaces of constant $r$ in the conformally regulated metric
(\ref{eq:deSreg}) is plotted over a range from the origin to beyond the
location $r=2$ of the surface which becomes Scri+ in the high resolution
limit $L \rightarrow \infty$. This example uses $L/s=10$.
}
\end{figure}

\begin{figure}
\includegraphics[scale=0.3, angle=-90]{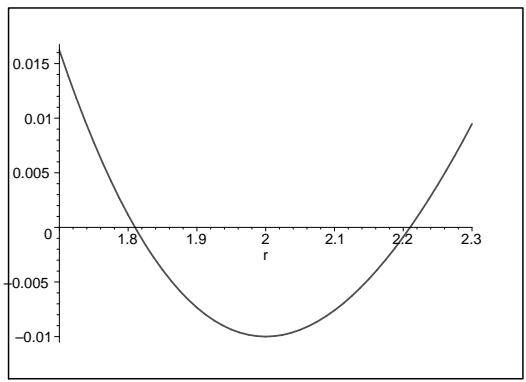}%
\caption{\label{fig:gradr2micro}
The squared norm $(\nabla r)^2 = g^{rr}$ of the normal $\nabla r$ to
hypersurfaces of constant $r$ in the conformally regulated metric
(\ref{eq:deSreg}) is plotted over a narrow range arround $r=2$ where $r
= \text{const}$ are spacelike pure-outflow hypersurfaces on which the
entire forward light cone is oriented toward increasing $r$ when $g^{rr}
< 0$. This example uses $L/s=10$.
}
\end{figure}

The causal structure of the regulated metric (\ref{eq:deSreg}) is
simpler than that of the original metric (\ref{eq:deSorig}). The
hypersurfaces of constant $u$ are everywhere spacelike since the
regulated metric is positive definite when $du =0$.
The hypersurfaces of constant $r$ are spacelike only in an interval
around $r=2$ as seen from their normal vector $dr$ being timelike there:
\begin{equation}
\label{eq:regnorm_dr} 
(\nabla r)^2 = g^{rr} =
  \frac{(1-\tfrac{1}{4}r^2)^2 - (rs/L)^2}{(1+\tfrac{1}{4}r^2)^2}
 \quad . 
\end{equation}
This norm is plotted in Figure~\ref{fig:gradr2macro} over a broad range
and again in Figure~\ref{fig:gradr2micro} giving the interval around $r=2$
in more detail.


\section{\label{sec:analysis}Analysis}
What can we learn using the tools described above? 
A few lessons already appeared in the work \cite{MSL:KITP} with Scheel
and Lindblom. Suppose only the time coordinate were changed to
parameterize a family of hyperboloids, here 
\begin{equation}
\label{eq:hyperboloid}
     [T-s(u-1)]^2 - R^2 = s^2
\end{equation}
in the original coordinates of equation~(\ref{eq:deSorig}). Then the
outgoing coordinate speed of light $dR/du$ would become unboundedly
large as $R \rightarrow \infty$. The Courant condition usually met in
numerical implementations would then make the allowable time step
impractically small. This can be cured by ``Analytic Mesh Refinement''
(AnMR), which is just a coordinate transformation such as that from $R$
to $r$ in equation~(\ref{eq:AnMR}), since outgoing waves get redshifted
as the $u=\text{const}$ hypersurfaces tend toward outgoing null cones.
Because this redshift gives long wavelengths, high resolution is not
needed and the large $R$ domain can be compressed to a finite range.
Thus we find that the outgoing coordinate speed of light $dr/du$ becomes
\begin{equation}
\label{eq:c_out}
                  c_\text{out} = (1 + \tfrac{1}{2} r)^2 
\end{equation}
which remains less than $4$ out to $r=2$ or $R=\infty$. The ingoing
radial light speed $c_\text{in}$ is similarly bounded in magnitude but
becomes positive in the region where $(\nabla r)^2 $ is negative rather
than the simpler value $c_\text{in} = - (1-\tfrac{1}{2} r)^2$ in the
Minkowski ($L=\infty$) case where it merely touches zero on Scri+. Thus
the encouraging results from \cite{MSL:KITP} are unchanged by
introducing the cosmological term.

\section{\label{sec:init}Exact Solutions}

There is in principle no difficulty in finding exact solutions to these
equations in Minkowski spacetime. One merely takes solutions from any
textbook and performs the coordinate transformation to our $uxyz$
compactified hyperboloidal slicing coordinates. In detail is it a great
convenience to have GRTensor \cite{grt} available to perform some of the
algebra and much essential checking.

Exact solutions play two roles. The first is that they provide initial
data for numerical integrations whose physical interpretation is known.
Secondly, though comparison, they allow testing the accuracy of the
numerical solution. Indeed, the whole point of carrying out these
numerical evolutions is to explore in a simpler context than full
general relativity different approaches for possible adaption to
Einstein's equations. Thus one may explore and compare different
integration schemes either to understand them better as in the work of
Baumgarte \cite{baumgarte:em}, or to test possible new schemes as in
\cite{fiske:constem}. Questions which may be treated in this simplified
context include the stability of different formulations, the control of
constraints, the treatment of boundary conditions, the extraction of
asymptotic wave amplitudes and waveforms, etc.  Here we address the use
of hyperboloidal time slices and the possibilities for carrying the
integration out to and beyond Scri+ using a finite grid.

\begin{figure*}
\includegraphics[scale=0.6, angle=-90]{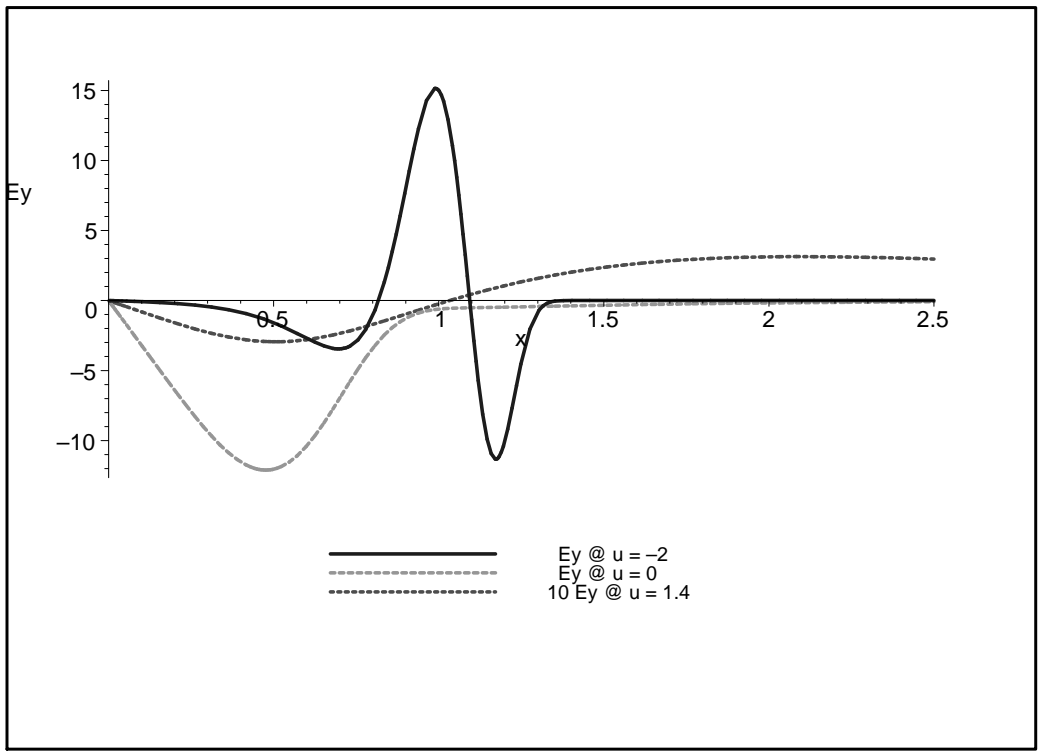}%
\caption{\label{fig:analytic}For the Baumgarte choice of EM wave packet
from equations (\ref{eq:Atemplate}) and (\ref{eq:Bwave}) we plot the
conformally invariant electric field $\mathfrak{D}^y$ along the $x$-axis at three
different hyperboloidal times $u = -2, 0, \text{and\ } 1.4$ . Note that
the field values are finite and generally nonzero at Scri+ which is
$x=2$. Those at the latest time have been here magnified by a factor of
10.}
\end{figure*}

We find that a sufficient variety of examples can be found by asuming
that the vector potential $A = A_\mu dx^\mu$ has the form
\begin{equation} 
\label{eq:Atemplate} 
           A = f \sin^2\theta\,d\phi -\psi\,dT \quad . 
\end{equation} 
Three cases are 
(1) a uniform magnetic field with $\psi=0, f =B_Z R^2/2$ 
(2) a static dipole electric field with $f=0, \psi = Z/R^3$ 
(3) the Baumgarte \cite{baumgarte:em} choice of a wave packet, which is
our principle test case, with $\psi=0$ and
\begin{equation} 
\label{eq:Bwave} 
       f =  (\frac{1}{R}-2 \lambda U)\exp(-\lambda U^2)
           -(\frac{1}{R}+2 \lambda V)\exp(-\lambda V^2)
\end{equation}
where $U$ and $V$ were defined by
\begin{subequations}
\label{eq:MinkUV}
\begin{equation}
  U \equiv T-R = s \left(u - \frac{r}{1+\tfrac{1}{2}r} \right)
\end{equation}
\text{and}
\begin{equation}
  V \equiv T+R = s \left(u + \frac{r}{1-\tfrac{1}{2}r} \right) \quad .
\end{equation}
\end{subequations}
In spite of first impressions one finds that $f = {\rm O}(R^2)$ at small
$R$ or small $r$, but numerical implementations must avoid evaluating 
this $f$ at $r=0$ as subtle cancellations occur.
To convert equation (\ref{eq:Atemplate}) to $uxyz$ coordinates requires
\begin{equation} 
\label{eq:dT} 
       \frac{1}{s} dT = du + \frac{x^i}{(1-\tfrac{1}{4}r^2)^2} dx^i
\end{equation}
and, from $X^i = s x^i/(1-\tfrac{1}{4}r^2)$ and the standard rectangular
to spherical coordinate transformation,
\begin{equation} 
\label{eq:dphi} 
       \sin^2\theta \,d\phi = \frac{1}{R^2}(X\,dY- Y\,dX)
                     = \frac{1}{r^2}(x\,dy- y\,dx)\ .
\end{equation}
The results are nontrivial even for a constant magnetic field in these
coordinates which can be useful for debugging. To obtain initial
conditions for numerical integrations requires that from these vector
potentials one compute the fields $F_{\mu\nu}$, $\mathfrak{F}^{\mu\nu}$
and thus $\mathfrak{D}^i = \mathfrak{F}^{0i}$. Figure~\ref{fig:analytic} plots an
analytic solution obtained in this way at three different times. The
incoming wave packet (at $u=-2$) is blueshifted relative to the normals
the to constant $u$ hyperboloid, while the outgoing packet (at $u=1.4$)
is redshifted so that high spatial resolution is not required to
describe it near Scri+ (which is at $r=2$). Note that there are no
irregularities of the outgoing packet as it crosses Scri+.

\begin{figure}
\includegraphics[scale=0.35, angle=-90]{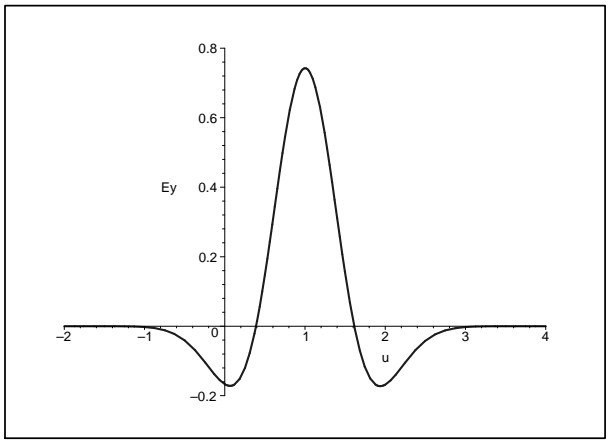}%
\caption{\label{fig:analWaveform}Waveform extraction: for the same
analytic solution as in Figure~\ref{fig:analytic} we plot $\mathfrak{D}^y$ as
function of $u$ at Scri+ ($x=2$). Note that at constant $r$ one has
$T=su + \text{const}$ so this waveform is not distorted from its $(T,R)$
presentation as is the pulse shape (Figure~\ref{fig:analytic}) when
plotted as function of $r$.}
\vspace*{7ex}
\end{figure}

%
%

\begin{acknowledgments}
This research was supported in part by the National Science Foundation
under Grant No.\ PHY99-07949 to UCSB and is an outgrowth of a project
begun during the KITP workshop ``Gravitational Interaction of Compact
Objects''. It was also supported in part by NSF Grant No.\ PHY00-71020 to
the University of Maryland, and NASA Grant ATP02-0043-0056.
\end{acknowledgments}



\bibliography{Misner}

\end{document}